\documentclass[conference]{IEEEtran}
\usepackage{subfigure}
\usepackage{amssymb}
\usepackage[english]{babel}
\usepackage{amsmath,bm}
\usepackage{color}
\usepackage{tabularx}
\usepackage{multirow}
\usepackage{array}

\newcommand{\tabincell}[2]{\begin{tabular}{@{}#1@{}}#2\end{tabular}}
\newtheorem{definition}{Definition}
\newtheorem{gp}{Guiding Principle}
\newtheorem{insight}{Insight}
\newcommand{\ignore}[1]{}

\usepackage{ifpdf}

\usepackage{cite}

\ifCLASSINFOpdf
   \usepackage[pdftex]{graphicx}
   \graphicspath{{../pdf/}{../jpeg/}}
   \DeclareGraphicsExtensions{.pdf,.jpeg,.png}
\else
   \usepackage[dvips]{graphicx}
   \graphicspath{{../eps/}}
   \DeclareGraphicsExtensions{.eps}
\fi
\usepackage{algorithmic}

\usepackage{array}
\usepackage{url}

\begin{document}
%
\title{VulDeePecker: A Deep Learning-Based System for Vulnerability Detection}



%

\author{\IEEEauthorblockN{Zhen Li\IEEEauthorrefmark{1}\IEEEauthorrefmark{2},
Deqing Zou\IEEEauthorrefmark{1}\IEEEauthorrefmark{3}$^\sharp$\thanks{$^\sharp$Corresponding author},
Shouhuai Xu\IEEEauthorrefmark{4},
Xinyu Ou\IEEEauthorrefmark{1},
Hai Jin\IEEEauthorrefmark{1},\\
Sujuan Wang\IEEEauthorrefmark{1},
Zhijun Deng\IEEEauthorrefmark{1} and
Yuyi Zhong\IEEEauthorrefmark{1}}
\IEEEauthorblockA{\IEEEauthorrefmark{1}Services Computing Technology and System Lab,
Big Data Technology and System Lab,\\
Cluster and Grid Computing Lab,
School of Computer Science and Technology,\\
Huazhong University of Science and Technology\\
deqingzou@hust.edu.cn}
\IEEEauthorblockA{\IEEEauthorrefmark{2}School of Cyber Security and Computer, Hebei University}
\IEEEauthorblockA{\IEEEauthorrefmark{3}Shenzhen Huazhong University of Science and Technology Research Institute}
\IEEEauthorblockA{\IEEEauthorrefmark{4}Department of Computer Science, University of Texas at San Antonio}
}


\IEEEoverridecommandlockouts
\makeatletter\def\@IEEEpubidpullup{9\baselineskip}\makeatother
\IEEEpubid{\parbox{\columnwidth}{
    Network and Distributed Systems Security (NDSS) Symposium 2018\\
    18-21 February 2018, San Diego, CA, USA\\
    ISBN 1-1891562-49-5\\
    http://dx.doi.org/10.14722/ndss.2018.23158\\
    www.ndss-symposium.org
}
\hspace{\columnsep}\makebox[\columnwidth]{}}

\maketitle

\begin{abstract}
The automatic detection of software vulnerabilities is an important research problem. However, existing solutions to this problem rely on human experts to define features and often miss many vulnerabilities (i.e., incurring high false negative rate). In this paper, we initiate the study of using deep learning-based vulnerability detection to relieve human experts from the tedious and subjective task of manually defining features. Since deep learning is motivated to deal with problems that are very different from the problem of vulnerability detection, we need some guiding principles for applying deep learning to vulnerability detection. In particular, we need to find representations of software programs that are suitable for deep learning. For this purpose, we propose using {\em code gadgets} to represent programs and then transform them into vectors, where a code gadget is a number of (not necessarily consecutive) lines of code that are semantically related to each other. This leads to the design and implementation of a deep learning-based vulnerability detection system, called \underline{Vul}nerability \underline{Dee}p \underline{Pecker} (VulDeePecker). In order to evaluate VulDeePecker, we present the first vulnerability dataset for deep learning approaches. Experimental results show that VulDeePecker can achieve much fewer false negatives (with reasonable false positives) than other approaches. We further apply VulDeePecker to 3 software products (namely Xen, Seamonkey, and Libav) and detect 4 vulnerabilities, which are not reported in the National Vulnerability Database but were ``silently'' patched by the vendors when releasing later versions of these products; in contrast, these vulnerabilities are almost entirely missed by the other vulnerability detection systems we experimented with.
\end{abstract}



%

\section{Introduction}
Many cyber attacks are rooted in software vulnerabilities.
Despite the effort that has been invested in pursuing secure programming, software vulnerabilities remain, and will continue, to be a significant problem. This can be justified by the fact that the number of vulnerabilities registered in the Common Vulnerabilities and Exposures (CVE) was approximately 4,600 in 2010, and grew to approximately 6,500 in 2016 \cite{CVE}. An alternate approach is to automatically detect vulnerabilities in software programs, or simply {\em programs} for short. There have been many static vulnerability detection systems and studies for this purpose, ranging from open source tools \cite{FlawFinder,RATS,ITS4}, to commercial tools \cite{Checkmarx,HP_Fortify,Coverity},
to academic research projects \cite{neuhaus2007predicting,shin2011evaluating,neuhaus2009beauty,yamaguchi2011vulnerability,yamaguchi2012generalized,grieco2016toward,li2016vulpecker,kim2017vuddy}.
However, existing solutions for detecting vulnerabilities have two major drawbacks: imposing {\em intense manual labor} and incurring {\em high false negative rates}, which are elaborated below.

On one hand, existing solutions for vulnerability detection rely on human experts to define features.
Even for experts, this is a tedious, subjective, and sometimes error-prone task because of the complexity of the problem. In other words, the identification of features is largely an art, meaning that the quality of the resulting features, and therefore the effectiveness of resulting detection system, varies with the individuals who define them.
In principle, this problem can be alleviated by asking multiple experts to define their own features,
and then select the set of features that lead to better effectiveness or use a combination of these features. However, this imposes even more tedious work.
As a matter of fact, it is always desirable to reduce, or even eliminate whenever possible, the reliance on the intense labor of human experts. This can be justified by the trend of cyber defense automation, which is catalyzed by initiatives such as DARPA's Cyber Grand Challenge \cite{CGC}.
It is therefore important to relieve human experts from the tedious and subjective task of manually defining features for vulnerability detection.

On the other hand, existing solutions often miss many vulnerabilities or incur high {\em false negative rates}.
For example, two most recent vulnerability detection systems, VUDDY \cite{kim2017vuddy} and VulPecker \cite{li2016vulpecker},
respectively incur a false negative rate of 18.2\% (when detecting vulnerabilities of Apache HTTPD 2.4.23) and 38\% (when applied to 455 vulnerability samples). Our own independent experiments show that they respectively incur a false negative rate of 95.1\% and 89.8\% (see Table \ref{Table_Comparison_with_other_tools} in Section \ref{sec:Experimental}).
Note that the large discrepancy between the false negative rates reported in \cite{kim2017vuddy,li2016vulpecker} and the false negative rates derived from our experiments is caused by the use of different datasets. These high false negative rates may be justified by their emphasis on low false positive rates, which are respectively 0\% for VUDDY \cite{kim2017vuddy} and unreported for VulPecker \cite{li2016vulpecker}. Our independent experiments show that their false positive rates are respectively 0\% for VUDDY and 1.9\% for VulPecker (see Table \ref{Table_Comparison_with_other_tools} in Section \ref{sec:Experimental}). This suggests that VUDDY and VulPecker are designed to achieve low false positive rates, which appear to be inherent to the approach of detecting vulnerabilities caused by code clones; in contrast, when using this approach to detecting vulnerabilities that are not caused by code clones, high false negative rates occur.

It would be fair to say that vulnerability detection systems with high false positive rates may not be {\em usable}, vulnerability detection systems with high false negative rates may not be {\em useful}. This justifies the importance of pursuing vulnerability detection systems that can achieve low false negative rates and low false positive rates. When this cannot be achieved (because false positive and false negative are often at odds with each other), we may put emphasis on lowering the false negative rate as long as the false positive rate is not too high.

The aforementioned two limitations of existing solutions motivate the importance of designing the vulnerability detection system {\em without} asking human experts to manually define features and {\em without} incurring high false negative rate or false positive rate. In this paper, we propose a solution to the following vulnerability detection problem while bearing in mind with these limitations:
{\em Given the source code of a target program, how can we determine whether or not the target program is vulnerable and if so, where are the vulnerabilities?}

\medskip
\noindent{\bf Our contributions.}
The present paper represents a first step towards ultimately tackling the aforesaid problem.
Specifically, we make three contributions.

First, we initiate the study of using deep learning for vulnerability detection. This approach has a great potential because deep learning does not need human experts to manually define features, meaning that vulnerability detection can be automated. However, this approach is challenging because deep learning is {\em not} invented for this kind of applications, meaning that we need some guiding principles for applying deep learning to vulnerability detection. We discuss some preliminary guiding principles for this purpose, including the representation of software programs to make deep learning suitable for vulnerability detection, the determination of {\em granularity} at which deep learning-based vulnerability detection should be conducted, and the selection of specific neural networks for vulnerability detection. In particular, we propose using {\em code gadgets} to represent programs. A code gadget is a number of (not necessarily consecutive) lines of code that are semantically related to each other, and can be vectorized as input to deep learning.

Second, we present the design and implementation of a deep learning-based vulnerability detection system, called \underline{Vul}nerability \underline{Dee}p \underline{Pecker} (VulDeePecker).
We evaluate the effectiveness of VulDeePecker from the following perspectives:
\begin{itemize}
\item Can VulDeePecker deal with multiple types of vulnerabilities at the same time?
This perspective is important because a target program in question may contain multiple types of vulnerabilities, meaning that a vulnerability detection system that can detect only one type of vulnerabilities would be too limited. Experimental results answer this question affirmatively. This can be explained by the fact that VulDeePecker uses vulnerability patterns (learned as deep neural networks) to detect vulnerabilities.

\item Can human expertise help improve the effectiveness of VulDeePecker?
Experimental results show that the effectiveness of VulDeePecker can be further improved by incorporating human expertise, which is not for defining features though.
This hints that automatic vulnerability detection systems, while being able to relieve human experts from the tedious labor of defining features,
may still need to leverage human expertise from other purposes. This poses an important open problem for future study.

\item How effective is VulDeePecker when compared with other vulnerability detection approaches?
Experimental results show that VulDeePecker is much more effective
than the other static analysis tools, which ask human experts to define rules for detecting vulnerabilities, and the state-of-the-art code similarity-based vulnerability detection systems (i.e., VUDDY and VulPecker).
\end{itemize}

These questions may be seen as an initial effort at defining a benchmark for evaluating the effectiveness of deep learning-based vulnerability detection systems.

In order to show the usefulness of VulDeePecker, we further apply it to 3 software products (namely Xen, Seamonkey, and Libav). VulDeePecker is able to detect 4 vulnerabilities,
which are not reported in the National Vulnerability Database (NVD) \cite{NVD} but were ``silently'' patched by the vendors when releasing later versions of these products.
In contrast, these vulnerabilities are almost entirely missed by the other vulnerability detection systems we experimented with. More precisely, one of those vulnerability detection systems is able to detect 1 of the 4 vulnerabilities (i.e., missing 3 of the 4 vulnerabilities), while the other systems missed all of the 4 vulnerabilities.
We will conduct more experiments to show whether or not VulDeePecker can detect vulnerabilities that have not been identified, including possibly 0-day vulnerabilities.

Third, since there are no readily available datasets for answering the questions mentioned above, we present the first dataset for evaluating VulDeePecker and other deep learning-based vulnerability detection systems that will be developed in the future. The dataset is derived from two data sources maintained by the National Institute of Standards and Technology (NIST): the NVD \cite{NVD} and the Software Assurance Reference Dataset (SARD) project \cite{SARD}. The dataset contains 61,638 code gadgets, including 17,725 code gadgets that are vulnerable and 43,913 code gadgets that are not vulnerable. Among the 17,725 code gadgets that are vulnerable, 10,440 code gadgets correspond to buffer error vulnerabilities (CWE-119) and the rest 7,285 code gadgets correspond to resource management error vulnerabilities (CWE-399). We have made the dataset available at \url{https://github.com/CGCL-codes/VulDeePecker}.

\medskip

\noindent{\bf Paper organization.}
The rest of the paper is organized as follows.
Section \ref{sec:Background} presents some preliminary guiding principles for deep learning-based vulnerability detection.
Section \ref{sec:Design} discusses the design of VulDeePecker.
Section \ref{sec:Experimental} describes our experimental evaluation of VulDeePecker and results.
Section \ref{sec:Limitations} discusses the limitations of VulDeePecker and open problems for future research.
Section \ref{sec:Related_work} describes the related prior work.
Section \ref{sec:Conclusion} concludes the present paper.

\section{Guiding Principles for Deep Learning-Based Vulnerability Detection}
\label{sec:Background}
In this section, we propose some preliminary guiding principles for using deep learning to detect vulnerabilities. These principles are sufficient for the present study, but may need to be refined to serve the more general purpose of deep learning-based vulnerability detection. These principles are centered at answering three fundamental questions:
(i) How to represent programs for deep learning-based vulnerability detection?
(ii) What is the appropriate granularity for deep learning-based vulnerability detection?
(iii) How to select a specific neural network for vulnerability detection?

\subsection{How to represent software programs?}
Since deep learning or neural networks take vectors as input, we need to represent programs as vectors that are semantically meaningful for vulnerability detection.
In other words, we need to encode programs into vectors that are the required input for deep learning.
Note that we cannot arbitrarily transform a program into vectors because the vectors need to preserve the semantic information of the program.
This suggests us to use some intermediate representation as a ``bridge'' between a program and its vector representation, which is the actual input to deep learning.
This leads to the following:

\begin{gp}
\label{gp:gp-1}
Programs can be first transformed into some intermediate representation that can preserve (some of) the semantic relationships between the programs' elements (e.g., data dependency and control dependency). Then, the intermediate representation can be transformed into a vector representation that is the actual input to neural networks.
\end{gp}

As we will elaborate later, Guiding Principle \ref{gp:gp-1} leads us to propose an intermediate representation dubbed {\em code gadget}.
The term {\em code gadget} is inspired by the term of {\em gadget} in the context of code-reuse attacks (see, e.g., \cite{davi2015code}),
because a code gadget is a small number of (not necessarily consecutive) lines of code.

\subsection{What is an appropriate granularity?}
Since it is desirable not only to detect whether a program is vulnerable or not, but also to pin down the locations of the vulnerabilities, a finer granularity should be used for deep learning-based vulnerability detection. This means that vulnerability detection should not be conducted at the program or function level, which are too coarse-grained because a program or function may have many lines of code and pinning down the locations of its vulnerability can be a difficult task by itself. This leads to:

\begin{gp}
\label{gp:gp-2}
In order to help pin down the locations of vulnerabilities, programs should be represented at a finer granularity than treating a program or a function as a unit.
\end{gp}

Indeed, the aforementioned {\em code gadget} representation leads to a fine-grained granularity for vulnerability detection because a code gadget often consists of a small number of lines of code. This means that the code gadget representation naturally satisfies Guiding Principle \ref{gp:gp-2}.

\subsection{How to select neural networks?}
Neural networks have been very successful in areas such as image processing, speech recognition, and natural language processing (e.g., \cite{krizhevsky2012imagenet,hinton2012deep,pennington2014glove}), which are different from vulnerability detection. This means that many neural networks may not be suitable for vulnerability detection, and that we need some principles to guide the selection of neural networks that are suitable for vulnerability detection. Our examination suggests the following:

\begin{gp}
\label{gp:gp-3}
Because whether or not a line of code contains a vulnerability may depend on the context, neural networks that can cope with {\em contexts} may be suitable for vulnerability detection.
\end{gp}

This principle suggests that neural networks for natural language processing may be suitable for vulnerability detection because context is also important in natural language processing \cite{montemagni1998augmenting}. Putting the notion of {\em context} into the setting of the present paper, we observe that the argument(s) of a program function call is often affected by earlier operations in the program and may also be affected by later operations in the program.

Since there are many neural networks for natural language processing, let us start with Recurrent Neural Networks (RNNs) \cite{vinyals2015show,su2017lattice}.
These neural networks are effective in coping with sequential data, and indeed have been used for program analysis (but not for vulnerability detection purposes) \cite{white2016deep,shin2015recognizing,white2015toward,gu2016deep}. However, RNNs suffer from the Vanishing Gradient (VG) problem, which can cause ineffective model training \cite{bengio2009learning}. Note that the VG problem is inherited by the Bidirectional variant of RNNs, called BRNNs \cite{schuster1997bidirectional}. We would prefer neural networks that do not suffer from the VG problem.

The VG problem can be addressed with the idea of {\em memory cells} into RNNs, including the Long Short-Term Memory (LSTM) cell and the Gated Recurrent Unit (GRU) cell \cite{hochreiter1997long, cho2014properties}. Since the GRU does not outperform the LSTM on language modeling \cite{jozefowicz2015empirical}, we select LSTM for vulnerability detection and defer its comparison with GRU to future work. However, even LSTM may be insufficient for vulnerability detection because it is {\em unidirectional} (i.e., from earlier LSTM cells to later LSTM cells). This is because the argument(s) of a program function call may be affected by earlier statements in the program and may be also affected by the later statements. This suggests that unidirectional LSTM may be insufficient and that we should use Bidirectional LSTM (BLSTM) for vulnerability detection.

\begin{figure}[!htbp]
\setlength{\belowcaptionskip}{-0.3cm}
\centering
\includegraphics[width=.4\textwidth]{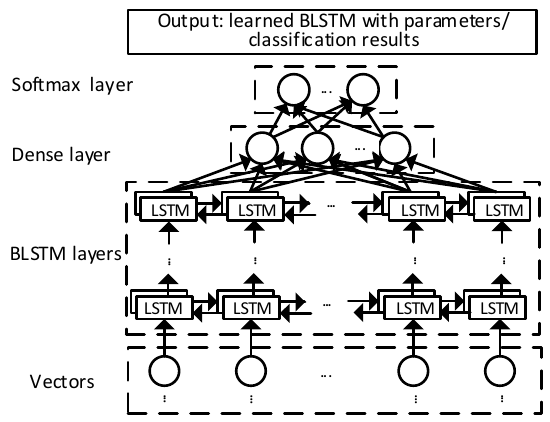}
\caption{A brief review of BLSTM neural network}
\label{Fig_blstm}
\end{figure}

Figure \ref{Fig_blstm} highlights the structure of BLSTM neural network, which has a number of {\em BLSTM} layers, a {\em dense} layer, and a {\em softmax} layer. The input to the learning process is in a certain vector representation. The BLSTM layers have two directions, {\em forward} and {\em backward}. The BLSTM layers contain some complex {\em LSTM cells}, which are treated as black-boxes in the present paper and therefore deferred to Appendix \ref{sec:appendix-LSTM-cells}. The dense layer reduces the number of dimensions of the vectors received from the BLSTM layers.
The softmax layer takes the low-dimension vectors received from the dense layer as input, and is responsible for representing and formatting the classification result, which provides feedback for updating the neural network parameters in the learning phase. The output of the learning phase is a BLSTM neural network with fine-tuned model parameters, and the output of the detection phase is the classification results.

\section{Design of VulDeePecker}
\label{sec:Design}
Our objective is to design a vulnerability detection system that can automatically tell whether a given program in source code is vulnerable or not and if so, the locations of the vulnerabilities. This should be achieved without asking human experts to manually define features and without incurring high false negative rates (as long as the false positive rates are reasonable). In this section, we describe the design of VulDeePecker. We start with a discussion on the notion of {\em code gadget}, because it is crucial to the representation of programs. Then, we give an overview of VulDeePecker and elaborate its components.

\subsection{Defining code gadget}

In order to represent programs in vectors that are suitable for the input to neural networks, we first propose transforming programs into a representation of {\em code gadget}, which is defined as follows:

\begin{definition}
\emph{(Code gadget)}
A {\em code gadget} is composed of a number of program statements (i.e., lines of code), which are semantically related to each other in terms of data dependency or control dependency.
\end{definition}

In order to generate code gadgets, we propose the heuristic concept of {\em key point}, which can be seen as a ``lens'' through which we can represent programs from a certain perspective.
Intuitively, the heuristic concept of {\em key point} can be seen as, in a sense, the ``center'' of a vulnerability or the piece of code that hints the existence of a vulnerability.
For vulnerabilities that are caused by improper uses of library/API function calls, the key points are the library/API function calls;
for vulnerabilities that are caused by improper uses of arrays, the key points are the arrays.
It is important to note that a type of vulnerabilities may have multiple kinds of key points.
For example, buffer error vulnerabilities may correspond to the following key points: library/API function calls, arrays, and pointers.
Moreover, the same kind of key points may exist in multiple types of vulnerabilities.
For example, both buffer error and resource management error vulnerabilities may contain the key points of library/API function calls.
Precisely defining the heuristic concept of key point is beyond the scope of the present paper and is left as an interesting problem for future research; instead, we focus on using this heuristic concept as the ``lens'' to use deep learning to learn vulnerability patterns.

In this paper, we focus on using the particular key point of library/API function calls to demonstrate its usefulness in deep learning-based vulnerability detection.
This is motivated by the observation that many vulnerabilities are related to library/API function calls.
It is also an interesting future work to investigate the usefulness of other kinds of key points.

Corresponding to the key point of library/API function calls, code gadgets can be generated by the means of data flow or control flow analysis of program, for which there are well known algorithms \cite{horwitz1990interprocedural,sinha1999system} and readily usable commercial products such as Checkmarx \cite{Checkmarx}.
It is worth mentioning that Checkmarx also detects vulnerabilities based on some rules that are {\em manually} defined by human experts.
However, we do not use its rules for vulnerability detection; instead, we will compare the effectiveness of VulDeePecker against it.

\subsection{Overview of VulDeePecker}

\begin{figure*}[!htb]
\centering
\subfigure[Learning phase]{\label{Fig_learning_phase}\includegraphics[width=.96\textwidth]{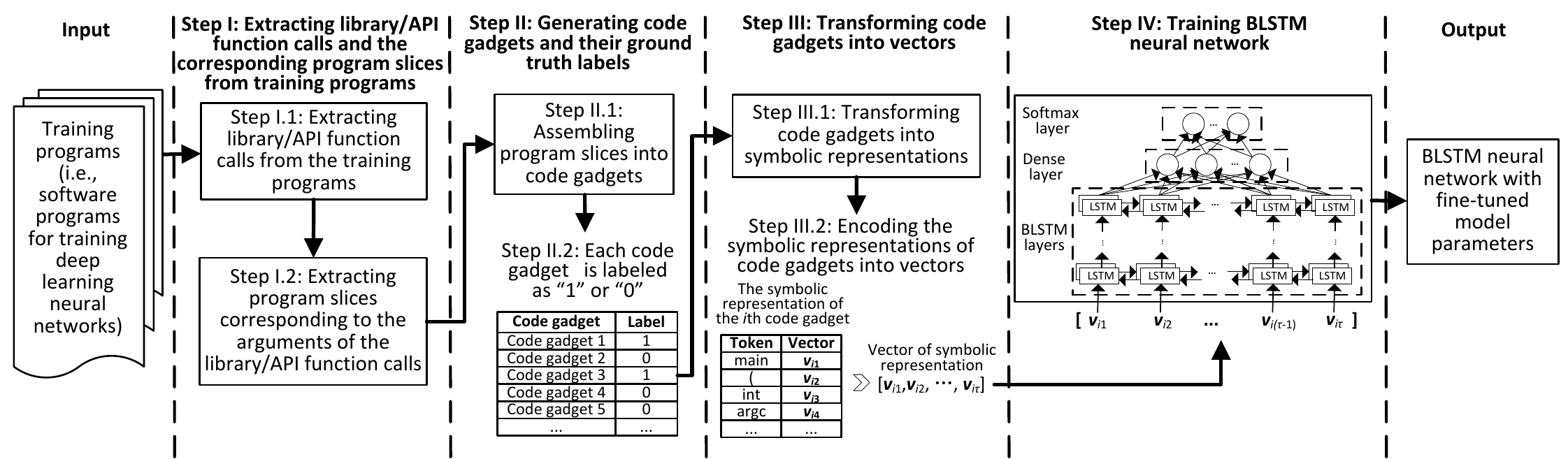}}
\subfigure[Detection phase]{\label{Fig_detection_phase}\includegraphics[width=.96\textwidth]{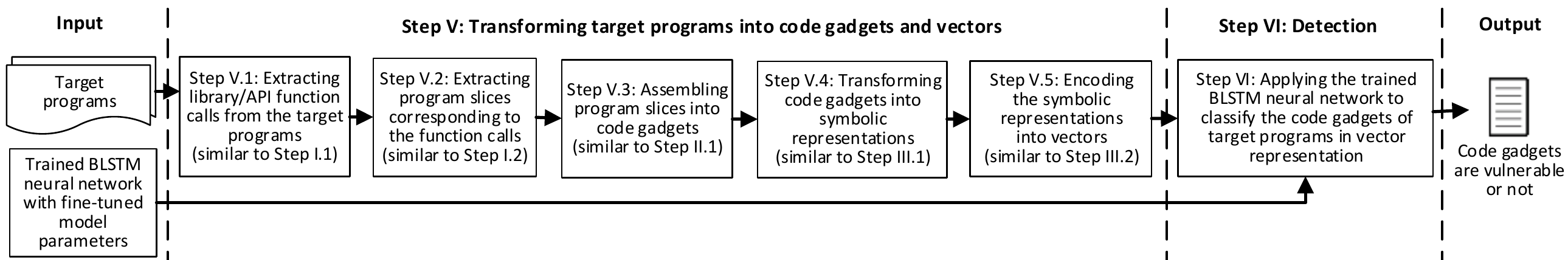}}
\caption{Overview of VulDeePecker: the learning phase generates vulnerability patterns, and the detection phase uses these vulnerability patterns to determine whether a target program is vulnerable or not and if so, the locations of the vulnerabilities (i.e., the corresponding {\em code gadgets}).}
\label{Fig_Overview_of_our_approach}
\end{figure*}

As highlighted in Figure \ref{Fig_Overview_of_our_approach}, VulDeePecker has two phases: a {\em learning} (i.e., training) phase and a {\em detection} phase. The input to the learning phase is a large number of {\em training programs},
some of which are vulnerable and the others are not. By ``vulnerable'' we mean that a program contains one or multiple known vulnerabilities.
The output of the learning phase is vulnerability patterns, which are coded into a BLSTM neural network.

\subsubsection{The learning phase}
As highlighted in Figure \ref{Fig_learning_phase}, the learning phase has 4 steps.

\noindent{\bf Step I: Extracting the library/API {\em function calls} and the corresponding {\em program slices}.}
This has two sub-steps, which are highlighted below and elaborated in Section \ref{sec:Step-I}.
\begin{itemize}
\item Step I.1: Extracting library/API function calls from the {\em training programs}, while noting that the current version of VulDeePecker focuses on vulnerabilities related to the key point of library/API function calls.
\item Step I.2: Extracting one or multiple program slices for each argument (or variable) of a library/API function call that is extracted in Step I.1.
In this paper, a program slice represents the statements of a program (i.e., lines of code) that are semantically related to an argument of a library/API function call, while noting that the notion of program slice was originally introduced to represent the statements of a program with respect to a program point or variable \cite{weiser1984program}.
\end{itemize}

\noindent{\bf Step II: Generating {\em code gadgets} of the training programs and their ground truth labels.}
 This step has two sub-steps, which are highlighted below and elaborated in Section \ref{sec:Step-II}.
\begin{itemize}
\item Step II.1: Assembling the program slices obtained in Step I.2 into code gadgets, one code gadget per library/API function call.
A code gadget does {\em not} necessarily correspond to some {\em consecutive} lines of code. Instead, it consists of multiple lines of code that are semantically related to each other (i.e., inheriting the semantic relation that is encoded in those program slices). 

\item Step II.2: Labeling the ground truth of code gadgets.
This step labels each code gadget as ``1'' (i.e., vulnerable) or ``0'' (i.e., not vulnerable).
The ground truth labels of code gadgets are available because we know whether a training program is vulnerable or not and if it is vulnerable, we also know the locations of the vulnerabilities.
\end{itemize}

\noindent{\bf Step III: Transforming code gadgets into vector representations.} This step has two sub-steps, which are highlighted below and elaborated in Section \ref{sec:Step-III}.
\begin{itemize}
\item Step III.1: Transforming code gadgets into certain symbolic representations, which will be elaborated later.
  This step aims to preserve some semantic information of the training programs.
\item Step III.2: Encoding code gadgets in the symbolic representation obtained in Step III.1 into vectors, which are the input for training a BLSTM neural network.
This is necessary in order to use neural networks in general.
\end{itemize}

\noindent{\bf Step IV: Training a BLSTM neural network.}
Having encoded the code gadgets into vectors and obtained their ground truth labels, the training process for learning a BLSTM neural network is standard.

\subsubsection{The detection phase}
Given one or multiple target programs, we extract library/API function calls from them  as well as the corresponding program slices, which are assembled into code gadgets.
The code gadgets are transformed into their symbolic representations, which are encoded into vectors and used as input to the trained BLSTM neural network. The network outputs which vectors, and therefore which code gadgets, are vulnerable (``1'') or not (``0'').
If a code gadget is vulnerable, it pins down the location of the vulnerability in the target program.
As highlighted in Figure \ref{Fig_detection_phase},
this phase has two steps.

\noindent{\bf Step V: Transforming target programs into code gadgets and vectors.} It has five sub-steps.
\begin{itemize}
\item Step V.1: Extracting library/API function calls from the target programs (similar to Step I.1).
\item Step V.2: Extracting program slices according to the arguments of the library/API function calls (similar to Step I.2).
\item Step V.3: Assembling the program slices into code gadgets (similar to Step II.1).
\item Step V.4: Transforming the code gadgets to their symbolic representations (similar to Step III.1).
\item Step V.5: Encoding the symbolic representations of code gadgets into vectors (similar to Step III.2).
\end{itemize}

\begin{figure*}[!htbp]
\centering
\includegraphics[width=.96\textwidth]{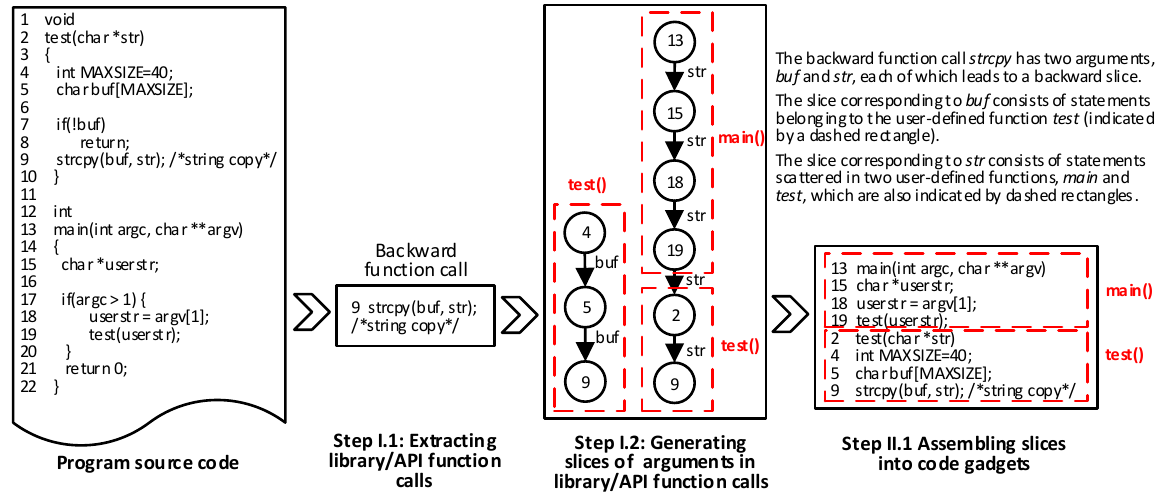}
\caption{Illustrating the extraction of library/API function calls (Step I.1) from a (training) program, which contains a backward function call (i.e., $strcpy$) that is also used as an example to demonstrate the extraction of program slices (Step I.2) and the assembly of program slices into code gadgets (Step II.1).}
\label{Fig_slice}
\end{figure*}

\noindent{\bf Step VI: Detection.} This step uses the learned BLSTM neural network to classify the vectors corresponding to the code gadgets that are extracted from the target programs.
When a vector is classified as ``1'' (i.e., vulnerable), it means that the corresponding code gadget is vulnerable
and the location of the vulnerability is pinned down. Otherwise, the corresponding code gadget is classified as ``0'' (i.e., not vulnerable).

\medskip

Steps I-III are respectively elaborated in the following subsections. Steps IV and VI are standard and Step V is similar to some of Steps I-III.

\subsection{Step I: Extracting library/API function calls and program slices}
\label{sec:Step-I}

\subsubsection{Step I.1: Extracting library/API function calls}
We classify library/API function calls into two categories:
{\em forward} library/API function calls and {\em backward} library/API function calls.
{\em Forward} library/API function calls are the function calls that receive one or multiple inputs
directly from the external input, such as the command line, a program, a socket, or a file.
For example, the $recv$ function call is a forward library/API function call because it receives data from a socket directly.
{\em Backward} library/API function calls are the function calls that do not receive any external input directly from the environment in which the program runs.

Figure \ref{Fig_slice} shows an example of
a {\em backward} library/API function call $strcpy$ (line 9). It is a {\em backward} library/API function call because it does not receive any external input directly.

We highlight a distinction between forward and backward library/API function calls. For forward library/API function calls, the statements influenced by the input arguments are critical because they may be vulnerable to improper (e.g., sophisticatedly crafted) argument values; for backward library/API function calls, the statements influencing the values of the arguments are critical because they could make the library/API function calls vulnerable. This insight will be leveraged to guide the heuristic padding of the vector representations of code gadgets.

\subsubsection{Step I.2: Extracting program slices}

This step generates program slices corresponding to the arguments of the library/API function calls that are extracted from the training programs.
We define two kinds of slices: {\em forward slices} and {\em backward slices}, where a forward slice corresponds to the statements that are affected by the argument in question and a backward slice corresponds to the statements that can affect the argument in question.
We take advantage of the commercial product Checkmarx \cite{Checkmarx}, more specifically its {\em data dependency graph}, to extract these two kinds of slices.
The basic idea is the following:
\begin{itemize}
\item For each argument in a {\em forward} library/API function call,
one or multiple {\em forward slices} are generated, with the latter corresponding to the case that the
slice related to the argument is {\em branched} at, or after, the library/API function call.
\item For each argument in a {\em backward} library/API function call, one or multiple {\em backward slices} are generated, with the latter corresponding to the case that multiple
    slices related to the argument are {\em merged} at, or prior to, the library/API function call.
\end{itemize}

Note that a program slice consists of multiple statements that may belong to multiple user-defined functions.
That is, a slice can go beyond the boundary of user-defined functions in question.

Figure \ref{Fig_slice} shows an example program that contains the library function call $strcpy$, which has two arguments $buf$ and $str$.
Since $strcpy$ is a backward function call, for each of its arguments we will generate a backward slice.
For argument $buf$, the slice consists of three statements, namely lines 4, 5, and 9 of the program, which belong to the user-defined function $test$.
For argument $str$, the slice consists of six statements, namely lines 13, 15, 18, 19, 2, and 9 of the program, where the first 4 belong to the user-defined function $main$
and the last 2 belong to the user-defined function $test$. The two slices are {\em chains} (i.e., a linear structure) because Checkmarx uses chains to represent slices, while noting that slices can also be represented by trees \cite{horwitz1990interprocedural,sinha1999system}. Since the linear structure can only represent one individual slice, a library/API function call often corresponds to multiple slices.

\subsection{Step II: Extracting code gadgets and labeling their ground truth}
\label{sec:Step-II}

\subsubsection{Step II.1: Assembling program slices into code gadgets}
\label{subsec:Extracting_code_fragments}
The program slices generated in the previous step can be assembled into code gadgets as follows.

First, given a library/API function call and the corresponding program slices, we combine the statements (i.e., pieces of code) belonging to the same, user-defined function into a single piece according to the order of the statements' appearance in the user-defined function. If there is a duplication of any statement, the duplication is eliminated.

In the example shown in Figure \ref{Fig_slice}, three statements (i.e., lines 4, 5, and 9) belonging to the user-defined function $test$ consists the program slice corresponding to the argument $buf$, and two statements (i.e., lines 2 and 9) belonging to the user-defined function $test$ are a piece of the program slice corresponding to the argument $str$. Therefore, we need to assemble them into a single piece because they are related to the same function $test$. According to the line numbers of these statements' appearance in the function $test$, this would lead to $2\rightarrow4\rightarrow 5\rightarrow 9 \rightarrow 9$. Since the statement corresponding to line 9 is duplicated, we eliminate the duplication to derive a piece of assembled statements $2\rightarrow4\rightarrow 5\rightarrow 9$, which correspond to the function $test$.

Second, assembling the statements belonging to different, user-defined functions into a single code gadget. If there is already an order between two pieces of statements belonging to these user-defined functions, this order is preserved; otherwise, a random order is used.

In the example shown in Figure \ref{Fig_slice}, when assembling the piece of statements belonging to the user-defined function $main$ (i.e., lines 13, 15, 18, and 19) and the assembled piece of statements belonging to user-defined function $test$ (i.e., lines 2, 4, 5, and 9), we obtain $13\rightarrow 15\rightarrow 18\rightarrow 19\rightarrow 2\rightarrow 4\rightarrow 5\rightarrow 9$, which is a {\em code gadget} corresponding to the library function call $strcpy$. This code gadget preserves the order of user-defined functions that are contained in the program slice corresponding to the argument $str$.

\subsubsection{Step II.2: Labeling the ground truth}
\label{subsec:Establishing_ground_truth}

Each code gadget needs to be labeled as ``1'' (i.e., vulnerable) and ``0'' (i.e., not vulnerable).
If a code gadget corresponds to a vulnerability that is known in the training dataset, it is labeled as ``1''; otherwise, it is labeled as ``0''.
In Section \ref{subsec:learning_BLSTM}, we will discuss the labeling of ground truth in details, when dealing with programs related to specific vulnerability data sources.

\subsection{Step III: Transforming code gadgets into vectors}
\label{sec:Step-III}

\subsubsection{Step III.1: Transforming code gadgets into their symbolic representations}

This step aims to heuristically capture some semantic information in the programs
for training a neural network.
First, remove the non-ASCII characters and comments because they have nothing to do with vulnerability.
Second, map user-defined variables to symbolic names (e.g., ``{\sf{VAR1}}'', ``{\sf{VAR2}}'') in the one-to-one fashion,
while noting that multiple variables may be mapped to the same symbolic name when they appear in different code gadgets.
Third, map user-defined functions to symbolic names (e.g., ``{\sf{FUN1}}'', ``{\sf{FUN2}}'') in the one-to-one fashion,
while noting that multiple functions may be mapped to the same symbolic name when they appear in different code gadgets.

\begin{figure}[!htbp]
\centering
\includegraphics[width=.48\textwidth]{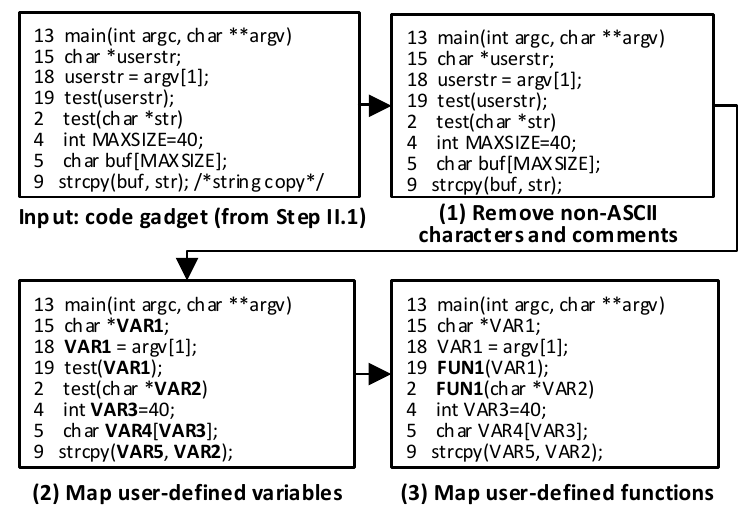}
\caption{Illustration of Step III.1: transforming code gadgets into their symbolic representations}
\label{Fig_identifier_mapping}
\end{figure}

Figure \ref{Fig_identifier_mapping} highlights the above process by using the code fragment generated by Step II.1 as shown in Figure \ref{Fig_slice}.

\subsubsection{Step III.2: Encoding the symbolic representations into vectors}

Each code gadget needs to be encoded into a vector via its symbolic representation. For this purpose, we divide a code gadget in the symbolic representation into a sequence of tokens via lexical analysis, including identifiers, keywords, operators, and symbols. For example, a code gadget in the symbolic representation,
$$``strcpy({\sf VAR5},~{\sf VAR2});"$$
is represented by a sequence of 7 tokens:
$$
``strcpy", ``{\sf{(}}", ``{\sf{VAR5}}", ``{\sf{,}}", ``{\sf{VAR2}}", ``{\sf{)}}", ~\text{and}~``{\sf{;}}".
$$
This leads to a large corpus of tokens. In order to transform these tokens into vectors, we use the {\em word2vec} tool \cite{word2vec}, which is selected because it is widely used in text mining \cite{wolf2014joint}. This tool is based on the idea of {\em distributed representation}, which maps a token to an integer that is then converted to a fixed-length vector \cite{rumelhart1986distributed}.

Since code gadgets may have different numbers of tokens, the corresponding vectors may have different lengths. Since BLSTM takes equal-length vectors as input, we need to make an adjustment. For this purpose, we introduce a parameter $\tau$ as the fixed length of vectors corresponding to code gadgets.
\begin{itemize}
\item When a vector is shorter than $\tau$, there are two cases: if the code gadget is generated from a {\em backward} slice or generated by combining multiple {\em backward} slices,
we pad zeros in the beginning of the vector; otherwise, we pad zeros to the end of the vector.
\item When a vector is longer than $\tau$, there are also two cases: if the code gadget is generated from one backward slice, or generated by combining multiple backward slices,
we delete the beginning part of the vector; otherwise, we delete the ending part of the vector.
\end{itemize}

This ensures that the last statement of every code gadget generated from a backward slice is a library/API function call, and that the first statement of every code gadget generated from a forward slice is a library/API function call.
As a result, every code gadget is represented
as a $\tau$-bit vector.
The length of vectors is related to the number of hidden nodes at each layer of the BLSTM, which is a parameter that can be tuned to improve the accuracy of vulnerability detection (see Section \ref{subsec:training_BLSTM}).

\section{Experiments and Results}
\label{sec:Experimental}
Our experiments are centered at answering the following three research questions (RQs):

\begin{itemize}
\item RQ1: Can VulDeePecker deal with multiple types of vulnerabilities at the same time?

A vulnerability detection system should be able to detect multiple types of vulnerabilities at the same time, because multiple detection systems need to be maintained otherwise.
For answering this question, we will conduct experiments involving one or multiple types of vulnerabilities.

\item RQ2: Can human expertise (other than defining features) improve the effectiveness of VulDeePecker?

For answering this question, we will investigate the effectiveness of using some manually-selected library/API function calls vs. the effectiveness of using all of the library/API function calls.

\item RQ3: How effective is VulDeePecker when compared with other vulnerability detection approaches?

For answering this question, we will compare VulDeePecker with other approaches,
including some static analysis tools and code similarity-based vulnerability detection systems.
\end{itemize}

\subsection{Metrics for evaluating vulnerability detection systems}
We use the widely used metrics {\em false positive rate} ($FPR$), {\em false negative rate} ($FNR$), {\em true positive rate} or {\em recall} ($TPR$), $precision$ ($P$),
and $F1\text{-}measure$ ($F1$) to evaluate vulnerability detection systems \cite{DBLP:journals/csur/PendletonGCX17}.
Let {\sf TP} be the number of samples with vulnerabilities detected correctly, {\sf FP} be the number of samples with false vulnerabilities detected, {\sf FN} be the number of samples with true vulnerabilities undetected, and {\sf TN} be the number of samples with no vulnerabilities undetected.
The false positive rate metric $FPR=\frac{{\sf FP}}{{\sf FP}+{\sf TN}}$ measures the ratio of false positive vulnerabilities to the entire population of samples that are not vulnerable.
The false negative rate metric $FNR=\frac{{\sf FN}}{{\sf TP}+{\sf FN}}$ measures the ratio of false negative vulnerabilities to the entire population of samples that are vulnerable.
The true positive rate or recall metric $TPR=\frac{{\sf TP}}{{\sf TP}+{\sf FN}}$ measures the ratio of true positive vulnerabilities   to the entire population of samples that are vulnerable,
while noting that $TPR=1-FNR$.
The precision metric $P=\frac{{\sf TP}}{{\sf TP}+{\sf FP}}$ measures the correctness of the detected vulnerabilities.
The $F1\text{-}measure$ metric $F1=\frac{2 \cdot P \cdot TPR }{P + TPR}$ takes consideration of both precision and true positive rate.

It would be ideal that a vulnerability detection system neither misses vulnerabilities (i.e., $FNR \approx 0$ and $TPR \approx 1$) nor triggers false alarms (i.e., $FPR \approx 0$ and $P \approx 1$), which means $F1 \approx 1$. However, this is difficult to achieve in practice, and often forces one to trade the effectiveness in terms of one metric for the effectiveness in terms of another metric. In this study, we prefer to achieving low FNR and low FPR.

\subsection{Preparing input to VulDeePecker}
\label{subsec:data_collection}
\noindent{\bf Collecting programs.}
There are two widely used sources of vulnerability data maintained by the NIST: the NVD \cite{NVD} which contains vulnerabilities in production software, and the SARD project \cite{SARD} which contains production, synthetic, and academic security flaws or vulnerabilities. In the NVD, each vulnerability has a unique Common Vulnerabilities and Exposures IDentifier (CVE ID) and a Common Weakness Enumeration IDentifier (CWE ID) that indicates the type of the vulnerability in question. We collect the programs that contain one or multiple vulnerabilities. In the SARD, each program (i.e., test case) corresponds to one or multiple CWE IDs because a program can have multiple types of vulnerabilities. Therefore, programs with one or multiple CWE IDs are collected.

In the present paper, we focus on two types of vulnerabilities: buffer error (i.e., CWE-119) and resource management error (i.e., CWE-399), each of which has multiple subtypes. These vulnerabilities are very common, meaning that we can collect enough data for using deep learning.
We select 19 popular C/C++ open source products, including the Linux kernel, Firefox, Thunderbird, Seamonkey, Firefox\_esr, Thunderbird\_esr,
Wireshark, FFmpeg, Apache Http Server, Xen, OpenSSL, Qemu, Libav, Asterisk, Cups, Freetype, Gnutls, Libvirt, and VLC media player, which contain, according to the NVD, these two types of vulnerabilities. We also collect the C/C++ programs in the SARD that contain these two types of vulnerabilities.
In total, we collect from the NVD 520 open source software programs related to buffer error vulnerabilities and 320 open source software programs related to resource management error vulnerabilities; we also collect from the SARD 8,122 programs (i.e., test cases) related to buffer error vulnerabilities and 1,729 programs related to resource management error vulnerabilities.
Note that program containing a vulnerability may actually consist of multiple program files.

\noindent{\bf Training programs vs. target programs.}
We randomly choose 80\% of the programs we collect as training programs and the remaining 20\% as target programs.
This ratio is applied equally when dealing with one or both types of vulnerabilities.

\subsection{Learning BLSTM neural networks}
\label{subsec:learning_BLSTM}

This corresponds to the learning phase of VulDeePecker.
We implement the BLSTM neural network in Python using Theano \cite{jamestheano} together with Keras \cite{Keras}.
We run experiments on a machine with NVIDIA GeForce GTX 1080 GPU and Intel Xeon E5-1620 CPU operating at 3.50GHz.

\noindent{\bf Step I: Extracting library/API function calls and corresponding program slices.}
We extract C/C++ library/API function calls from the programs.
There are 6,045 C/C++ library/API function calls that involve standard library function calls \cite{C_C++_library_functions}, basic Windows API and Linux kernel API function calls \cite{Windows_API_functions, Linux_kernel_API_functions}.
In total, we extract 56,902 library/API function calls from the programs,
including 7,255 {\em forward} function calls and 49,647 {\em backward} function calls.

In order to answer the RQs, we also {\em manually} select 124 C/C++ library/API function calls (including function calls with wildcard) related to buffer error vulnerabilities (CWE-119) and 16 C/C++ library/API function calls related to resource management error vulnerabilities (CWE-399). These function calls are selected because the aforementioned commercial tool Checkmarx \cite{Checkmarx} claims, using their own rules written by human experts, that they are related to these two types of vulnerabilities.
The list of these function calls are deferred to Table \ref{Table_sensitive_function} in the Appendix \ref{sec:appendix:library-and-API-functions}. Correspondingly, we extract 40,351 library/API function calls from the training programs, including 4,012 {\em forward} function calls and 36,339 {\em backward} function calls.
For each argument of the library/API function calls, one or multiple program slices are extracted.

\noindent{\bf Step II.1: Generating code gadgets.}
Code gadgets are generated from program slices. We obtain a Code Gadget Database (CGD) of 61,638 code gadgets,
among which 48,744 code gadgets are generated from program slices of training programs, and 12,894 code gadgets are generated from program slices of target programs.
The time complexity for generating gadgets mainly depends on the data flow analysis tool. For example, it takes 883 seconds to generate 2,494 code gadgets from 100 programs (99,232 lines) that are randomly selected the SARD, meaning an average of 354 milliseconds per code gadget.
For answering the RQs mentioned above, we use the CGD to derive the following 6 datasets.

\begin{itemize}
  \item BE-ALL: The subset of CGD corresponding to Buffer Error vulnerabilities (CWE-119) and ALL library/API function calls
   (i.e., extracted without human expert).
  \item RM-ALL: The subset of CGD corresponding to Resource Management error vulnerabilities (CWE-399) and ALL library/API function calls.
  \item HY-ALL: The subset of CGD corresponding to the HYbrid of (i.e., both) buffer error vulnerabilities (CWE-119) and resource management error vulnerabilities (CWE-399) and ALL library/API function calls. That is, it is the same as the CGD.
  \item BE-SEL: The subset of CGD corresponding to Buffer Error vulnerabilities (CWE-119) and manually SELected function calls (rather than all function calls).
  \item RM-SEL: The subset of CGD corresponding to Resource Management error vulnerabilities (CWE-399) and manually SELected function calls.
  \item HY-SEL: The subset of CGD corresponding to the HYbrid of buffer error vulnerabilities (CWE-119) and resource management error vulnerabilities (CWE-399) and manually SELected function calls.
\end{itemize}

\begin{table}[!htbp]
\caption{Datasets for answering the RQs}
\label{Table_dataset}
\centering
\begin{tabular}{|c|c|c|c|}
\hline
Dataset & \tabincell{c}{\#Code \\gadgets} & \tabincell{c}{\#Vulnerable \\code gadgets} & \tabincell{c}{\#Not vulnerable \\code gadgets}\\
\hline
BE-ALL &  39,753 & 10,440 & 29,313\\
\hline
RM-ALL & 21,885 & 7,285 & 14,600\\
\hline
HY-ALL &  61,638 & 17,725 & 43,913\\
\hline
BE-SEL &  26,720 & 8,119 & 18,601\\
\hline
RM-SEL & 16,198 & 6,573 & 9,625\\
\hline
HY-SEL & 42,918 & 14,692 & 28,226\\
\hline
\end{tabular}
\end{table}

Table \ref{Table_dataset} summarizes the number of code gadgets in these datasets.

\noindent{\bf Step II.2: Labeling code gadgets.}
Code gadgets are labeled as follows. For the code gadgets extracted from the programs of the NVD, we focus on the vulnerabilities whose patches involve line deletions or modifications. This process has two steps. In the first step, a code gadget is {\em automatically} labeled as ``1'' (i.e., vulnerable) if it contains at least one statement that is deleted or modified according to the patch, and labeled as ``0'' otherwise (i.e., not vulnerable). 
However, this automatic process may mislabel some code gadgets, which are not vulnerable, as ``1''. In order to remove these mislabels, the second step is to {\em manually} check the code gadgets that are labeled as ``1'' so as to correct the mislabels (if any).

Recall that each program in the SARD 
is already labeled as {\em good} (i.e., no security defect), {\em bad} (i.e., containing security defects), or {\em mixed} (i.e., containing functions with security defects and their patched versions) with corresponding CWE IDs.
For the code gadgets extracted from the programs with respect to the SARD,
a code gadget extracted from a {\em good} program is labeled as ``0'' (i.e., not vulnerable), and
a code gadget extracted from a {\em bad} or {\em mixed} program is labeled as ``1'' (i.e., vulnerable) if it contains at least one vulnerable statement and ``0'' otherwise.
Since we used heuristics in the labeling process for the SARD program,
we looked at the labels of 1,000 random code gadgets and found that only 6 of them (i.e. 0.6\%) were mislabeled.
These mislabeled samples are caused by the fact that a statement in a piece of code that is not vulnerable is the same as a statement in a piece of code that is vulnerable.
As the mislabeled code gadgets are very few and the neural networks are robust against a small portion of mislabeled samples, it is unnecessary to check manually all labels of the code gadgets that are extracted for the SARD programs.

It is possible to encounter the situation that the same code gadget is labeled with both ``1'' and ``0'' (i.e., conflicting labels).
One cause of this phenomenon is the imperfection of the data flow analysis tool.
In this case, we simply delete these code gadgets.
As a result, 17,725 code gadgets are labeled as ``1'' and 43,913 code gadgets are labeled as ``0''.
Among the 17,725 code gadgets labeled as ``1'', 10,440 code gadgets correspond to the buffer error vulnerabilities and 7,285 code gadgets correspond to the resource management error vulnerabilities.
Table \ref{Table_dataset} shows the number of code gadgets that are vulnerable (Column 3) and the number of code gadgets that are not vulnerable in each dataset (Column 4).

\noindent{\bf Step III: Transforming code gadgets into vectors.}
The CGD contains a total number of 6,166,401 tokens, of which 23,464 are different.
After mapping user-defined variable names and function names to some symbolic names, the number of different tokens is further reduced to 10,480.
These symbolic representations are encoded into vectors, which are used as the input for training a BLSTM neural network.

\noindent{\bf Step IV: Training BLSTM neural network.}
\label{subsec:training_BLSTM}
For each dataset described in Table \ref{Table_dataset}, we adopt a 10-fold cross validation to train a BLSTM neural network, and select the best parameter values corresponding to the effectiveness for vulnerability detection. For example, we vary the number of hidden layers for each BLSTM neural network and observe the influence on the resulting F1-measure.
When we adjust the number of hidden layers, we set the parameters to their default values when such default values are available,
and set the parameters to the values that are widely used by the deep learning community otherwise.
The number of tokens in the vector representation of code gadgets is set to 50,
the dropout is set to 0.5, the batch size is set to 64, the number of epochs is set to 4, the minibatch stochastic gradient descent together with ADAMAX \cite{kingma2014adam} is used for training with the default learning rate of 1.0, and 300 hidden nodes are chosen.

\begin{figure}[htbp!]
\centering
\includegraphics[width=.4\textwidth]{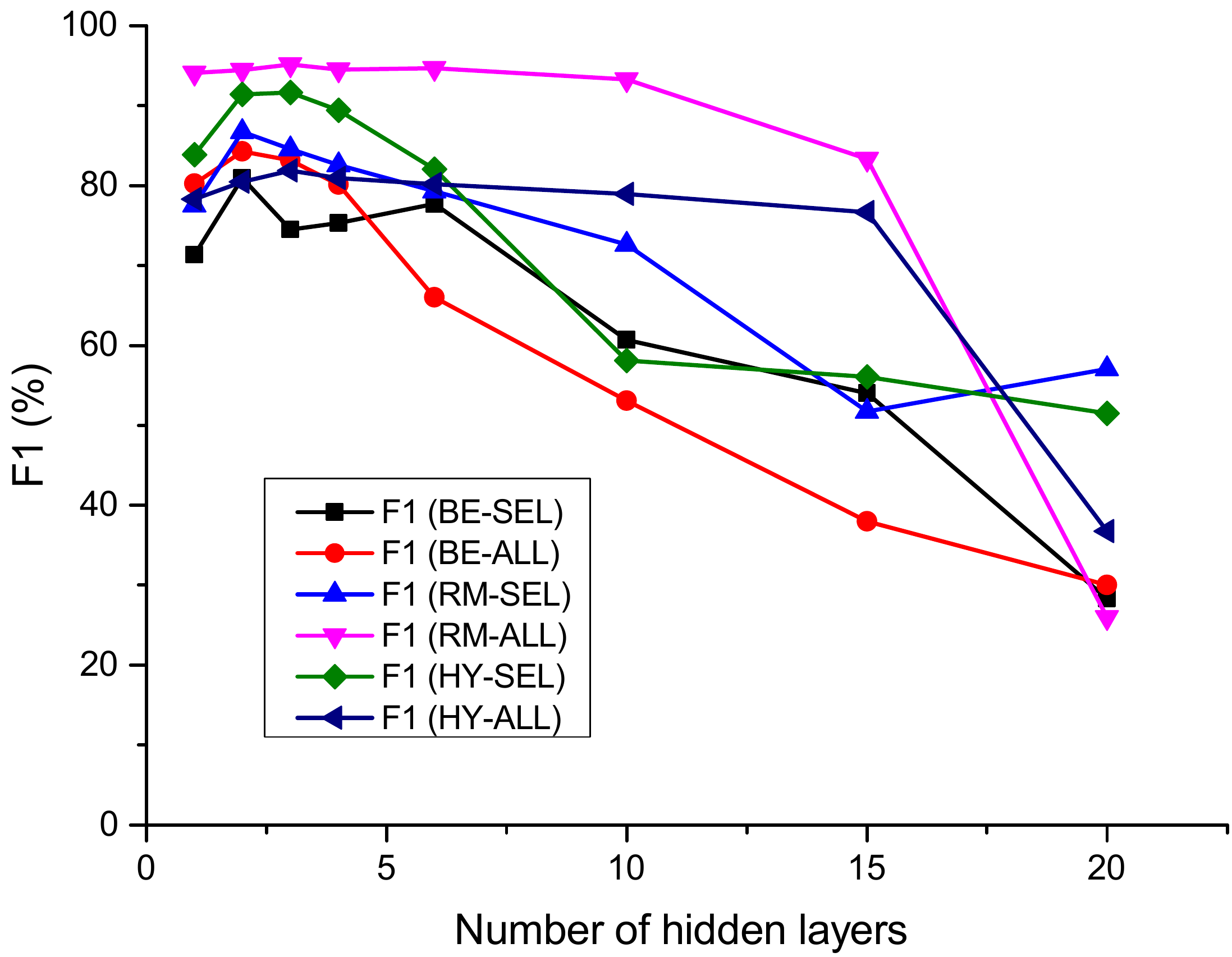}
\caption{F1-measure of VulDeePecker for the 6 datasets with different number of hidden layers}
\label{Fig_number_of_hidden_layers}
\end{figure}

Figure \ref{Fig_number_of_hidden_layers} plots the F1-measure of VulDeePecker with respect to the 6 datasets with different number of hidden layers, each of which leads to a different neural network. We observe that the F1-measure of the 6 BLSTM neural networks reaches the maximum at 2 or 3 layers, and the F1-measure of most of these BLSTM neural networks declines when the number of layers is greater than 6. Note that the other parameters of the BLSTM neural network can be tuned in a similar fashion.

\subsection{Experimental results \& implications}

\subsubsection{Experiments for answering RQ1}
In order to test whether VulDeePecker can be applied to multiple types of vulnerabilities,
we conduct experiments on three datasets: BE-ALL, RM-ALL, and HY-ALL.
This respectively leads to three neural networks, whose effectiveness is reported in Table \ref{Table_Comparison_vulnerability_types-for-RQ1}.

\begin{table}[!htbp]
\caption{Results for answering RQ1, indicating that VulDeePecker can detect multiple types of vulnerabilities.}
\label{Table_Comparison_vulnerability_types-for-RQ1}
\centering
\begin{tabular}{|c|c|c|c|c|c|}
\hline
Dataset & FPR(\%)& FNR(\%) & TPR(\%)& P(\%) & F1(\%) \\
\hline
BE-ALL & 2.9 & 18.0 & 82.0& 91.7 & 86.6\\
\hline
RM-ALL & 2.8 & 4.7 & 95.3& 94.6 & 95.0 \\
\hline
HY-ALL & 5.1 & 16.1 & 83.9& 86.9 & 85.4\\
\hline
\end{tabular}
\end{table}

We observe that the neural network trained from the RM-ALL dataset outperforms the neural network trained from the BE-ALL dataset in terms of all five metrics.
This can be explained by the fact that the number of library/API function calls that are related to resource management error vulnerabilities (i.e., 16) is far smaller than the number of library/API function calls that are related to buffer error vulnerabilities (i.e., 124). We also observe that, in terms of the FPR and P metrics, the neural network trained from the HY-ALL dataset is not as good as the neural network trained from the BE-ALL or RM-ALL dataset. We further observe that the TPR and FNR of the neural network trained from the HY-ALL dataset reside in between that of the neural network trained from the RM-ALL dataset and that of the neural network trained from the BE-ALL dataset. The F1-measure of the neural network trained from the HY-ALL dataset is 1.2\% lower than that of the neural network trained from the BE-ALL dataset, and 9.6\% lower than that of the neural network trained from the RM-ALL dataset. This can be explained by the fact that the number of library/API function calls that are related to the vulnerabilities of the hybrid dataset (i.e., 140) is larger than the number of library/API function calls that are related to a single type of vulnerabilities. We speculate this is caused by the following: it is more difficult to extract vulnerability patterns for a large number of library/API function calls that are related to vulnerabilities
than to extract vulnerability patterns for a small number of library/API function calls that are related to vulnerabilities.

Table \ref{Table_Performance_analysis_ALL} summarizes the training time and detection time corresponding to the HY-ALL dataset, where the second column represents the number of code gadgets for training (i.e., extracted from the training programs) and the third column represents the number of code gadgets for detection (i.e., extracted from the target programs).
We observe that the training time of VulDeePecker, as implied by the deep learning technology in general, is large, but the detecting time is small.

\begin{table}[!htbp]
\renewcommand{\arraystretch}{1.2}
\caption{Time complexity of training and detection}
\label{Table_Performance_analysis_ALL}
\centering
\begin{tabular}{|c|c|c|c|c|}
\hline
Dataset & \tabincell{c}{\#Training \\code gadgets} & \tabincell{c}{\#Detection \\code gadgets} & \tabincell{c}{Training \\time (s)}& \tabincell{c}{Detection \\time (s)}  \\
\hline
HY-ALL & 48,744 & 12,894 & 36,372.2 & 156.2  \\
\hline
HY-SEL &33,813 & 9,105 & 25,242.3 & 76.6  \\
\hline
\end{tabular}
\end{table}

In summary, we answer RQ1 affirmatively with the following:
\begin{insight}
VulDeePecker can simultaneously detect multiple types of vulnerabilities, but the effectiveness is sensitive to the number of library/API function calls related to vulnerabilities (i.e., the fewer the better).
\end{insight}

\subsubsection{Experiments for answering RQ2}
\label{subsubsec:Experiments_for_RQ2}
In order to answer whether VulDeePecker can be improved by incorporating human expertise,
we conduct the experiment using {\em all} library/API
function calls that are automatically extracted vs. using {\em some} library/API
function calls that are manually selected under the guidance of vulnerability rules written by Checkmarx's human experts.

\begin{table}[!htbp]
\caption{Results for answering RQ2, indicating that using manually-selected library/API function calls can indeed improve the effectiveness of VulDeePecker.}
\label{Table_Comparison_vulnerability_types}
\centering
\begin{tabular}{|c|c|c|c|c|c|}
\hline
Dataset & FPR(\%)& FNR(\%) & TPR(\%)& P(\%) & F1(\%) \\
\hline
HY-ALL & 5.1 & 16.1 & 83.9& 86.9 & 85.4\\
\hline
HY-SEL & 4.9 & 6.1 & 93.9& 91.9 & 92.9 \\
\hline
\end{tabular}
\end{table}

As shown in Table \ref{Table_Comparison_vulnerability_types}, the BLSTM network trained from the HY-SEL dataset is more effective than the BLSTM network trained from the HY-ALL dataset.
Although the improvement in FPR is small (0.2\%), the improvement in each of the other metrics is substantial: 10\% in FNR and TPR, 5\% in precision, and 7.5\% in F1-measure.
Moreover, Table \ref{Table_Performance_analysis_ALL} shows that the training time of using manually selected library/API function calls can be smaller than that of using all library/API function calls, because a smaller number of code gadgets need to be processed. This leads to the following preliminary understanding regards the usefulness of human expertise in improving the effectiveness of deep learning-based vulnerability detection:

\begin{insight}
Human expertise can be used to select library/API function calls to improve the effectiveness of VulDeePecker, especially the overall effectiveness in F1-measure.
\end{insight}

\subsubsection{Experiments for answering RQ3}
In order to answer RQ3, we compare the effectiveness of VulDeePecker
with other pattern-based and code similarity-based vulnerability detection systems.
We here report the comparison between their effectiveness in detecting buffer error vulnerabilities (i.e., BE-SEL dataset),
while noting that a similar phenomenon is observed when comparing their effectiveness in detecting resource management error vulnerabilities (i.e., RM-SEL dataset) --- the details are omitted due to the lack of space.

For comparison with other pattern-based vulnerability detection systems, which use rules defined by human experts, we consider
a commercial product called Checkmarx \cite{Checkmarx} and two open source tools called Flawfinder \cite{FlawFinder} and RATS \cite{RATS}.
These systems are chosen because we have access to them and to the best of our knowledge, they have been widely used.
For comparison with code similarity-based vulnerability detection systems, which are mainly geared towards clone-caused vulnerabilities,
we consider the two state-of-the-art systems called VUDDY \cite{kim2017vuddy} and VulPecker \cite{li2016vulpecker}.
We use VUDDY's open service, and use the original implementation of VulPecker provided to us by its authors.
For fair comparison, we need to address some subtle issues.
We observe that VulPecker uses diffs as an input, where a diff describes the difference between a vulnerable piece of code and its patched version,
we divide the BE-SEL dataset of target programs into two subsets, namely BE-SEL-NVD (266 samples derived from the NVD) and BE-SEL-SARD (the rest samples derived from the SARD). We use BE-SEL-NVD for the comparison study because VUDDY and VulPecker were designed to detect vulnerabilities with CVE IDs or vulnerabilities with diffs, but are unable to detect vulnerabilities in BE-SEL-SARD.

\begin{table}[!htbp]
\caption{Results for answering RQ3: VulDeePecker achieves a much smaller overall FNR of 7.0\% corresponding to the entire BE-SEL dataset (the larger FNR of 16.9\% corresponding to the small sub-dataset derived from the NVD and the smaller FNR of 5.1\% corresponding to the sub-dataset derived from the SARD), while noting that the overall FPR of 5.7\% is reasonably small. ``N/C'' means that the system is \underline{N}ot \underline{C}apable of detecting vulnerabilities in the corresponding dataset.}
\label{Table_Comparison_with_other_tools}
\centering
\begin{tabular}{|c|c|c|c|c|c|c|c|}
\hline
System & Dataset & \tabincell{c}{FPR\\(\%)} & \tabincell{c}{FNR\\(\%)} & \tabincell{c}{TPR\\(\%)}& \tabincell{c}{P\\(\%)} & \tabincell{c}{F1\\(\%)} \\
\hline
\multicolumn{7}{|c|}{VulDeePecker vs. Other pattern-based vulnerability detection systems} \\
\hline
Flawfinder & BE-SEL & 44.7 & 69.0 &  31.0& 25.0 & 27.7\\
\hline
RATS  & BE-SEL & 42.2 & 78.9 &  21.1& 19.4 & 20.2 \\
\hline
Checkmarx & BE-SEL & 43.1 & 41.1 & 58.9& 39.6 & 47.3\\
\hline
VulDeePecker & BE-SEL & {\bf 5.7} & {\bf 7.0} & 93.0& 88.1 & 90.5\\
\hline
\hline
\multicolumn{7}{|c|}{VulDeePecker vs. Code similarity-based vulnerability detection systems} \\
\hline
VUDDY & BE-SEL-NVD & 0 & 95.1 & 4.9& 100 & 9.3\\
\hline
VulPecker & BE-SEL-NVD & 1.9 & 89.8 & 10.2& 84.3 & 18.2\\
\hline
VulDeePecker & BE-SEL-NVD & {\bf 22.9} & {\bf 16.9} & 83.1& 78.6 &80.8\\
\hline
\hline
VUDDY        & BE-SEL-SARD & N/C & N/C & N/C& N/C & N/C\\
\hline
VulPecker    & BE-SEL-SARD & N/C & N/C & N/C& N/C & N/C\\
\hline
VulDeePecker & BE-SEL-SARD & {\bf 3.4} & {\bf 5.1} & 94.9 & 92.0 &93.4\\
\hline
\end{tabular}
\end{table}

Table \ref{Table_Comparison_with_other_tools} summarizes the comparison. We make the following observations.
First, VulDeePecker substantially outperforms the other pattern-based vulnerability detection systems, because VulDeePecker incurs a FPR of 5.7\% and a FNR of 7.0\%,
which are respectively much smaller than their counterparts in the other detection systems.
By looking into the other systems, we find that Flawfinder and RATS do not use data flow analysis and therefore miss many vulnerabilities.
Although Checkmarx does use data flow analysis, its rules for recognizing vulnerabilities are defined by human experts and are far from perfect. This further highlights the importance of
relieving human experts from tedious tasks (similar to task of manually defining features).
This observation leads to:

\begin{insight}
\label{insight:data-flow-significance}
A deep learning-based vulnerability detection system can be more effective by taking advantage of the data flow analysis. (This hints us to speculate that a system can be even more effective by taking advantage of the control flow analysis. It is an interesting future work to validate or invalidate this speculation.)
\end{insight}

Second, for the BE-SEL-NVD sub-dataset, VUDDY and VulPecker trade high FNRs (95.1\% and 89.8\%, respectively) for low FPRs (0\% and 1.9\%, respectively),
which lead to very low F1-measures (9.3\% and 18.2\%, respectively).
The large FNRs can explained by the following facts:
VUDDY can only detect the vulnerabilities related to functions, which are nearly identical to the vulnerable functions in the training programs (i.e., vulnerabilities caused by Types I and II code clones \cite{rattan2013software});
VulPecker can only detect vulnerabilities caused by Type I, Type II, and some Type III code clones \cite{rattan2013software} (e.g., deletion, insertion, and rearrangement of statements), which explains why VulPecker incurs a lower FNR than VUDDY.
However, these systems cannot detect vulnerabilities that are not caused by code clones, which explains why they incur high FNRs.

\begin{table*}[!htbp]
\renewcommand{\arraystretch}{1.2}
\caption{VulDeePecker detected 4 vulnerabilities in 3 products, which are not published in the NVD but have been ``silently'' patched by the vendors in the later releases of these products.These vulnerabilities are entirely missed by the other vulnerability detection systems,
except that FlawFinder detected only one vulnerability while missing the other three.}
\label{Vulnerabilities detected by VulDeePecker}
\centering
\begin{tabular}{|c|c|c|c|c|c|c|}
\hline
Target product & CVE ID & \tabincell{c}{Vulnerable product \\published in the NVD} & \tabincell{c}{Vulnerability \\publish time} & Vulnerable file in target product & \tabincell{c}{Library/API \\function call} & \tabincell{c}{1st patched version \\of target product} \\
\hline
{Xen 4.6.0} & CVE-2016-9104 & Qemu & 12/09/2016 &.../qemu-xen/hw/9pfs/virtio-9p.c & memcpy & Xen 4.9.0 \\
\hline
{Seamonkey} & CVE-2015-4517 & Firefox & 09/24/2015 & .../system/gonk/NetworkUtils.cpp & snprintf & Seamonkey 2.38 \\
\cline{2-7}
{2.31} & CVE-2015-4513 & Firefox & 11/05/2015 & .../protocol/http/Http2Stream.cpp & memset & Seamonkey 2.39 \\
\hline
{Libav 10.2} & CVE-2014-2263 & FFmpeg & 02/28/2014 & libavformat/mpegtsenc.c & strchr, strlen & Libav 10.4 \\
\hline
\end{tabular}
\end{table*}

In contrast, VulDeePecker has a much higher F1-measure (i.e., 80.8\% vs. 9.3\% for VUDDY and 18.2\% for VulPecker) because it has a much higher TPR (i.e., much lower FNR),
while noting that its FPR is 22.9\% (vs. 0\% for VUDDY and 1.9\% for VulPecker).
We suspect that this high FPR of 22.9\% corresponding to the BE-SEL-NVD dataset is caused by 
a small number of training code gadgets from NVD.
This can be justified by the small FPR of 3.4\% corresponding to 
a large number of training code gadgets from SARD, which is about 18 times larger than the number of training code gadgets from NVD.
Moreover, the FPR of 5.7\% corresponding to the entire BE-SEL dataset resides some where in between them.
The high FNR of 16.9\% can be explained similarly.

The high FPR and FNR of VulDeePecker with respect to the BE-SEL-NVD sub-dataset should not be used as evidence against VulDeePecker, simply because for the BE-SEL-SARD sub-dataset, VulDeePecker incurs an even smaller FPR of 3.4\% and a FNR of 5.1\%,
while noting that VUDDY and VulPecker are not applicable (i.e., not capable of detecting vulnerabilities in this sub-dataset).
Moreover, VulDeePecker incurs a FPR of 5.7\% and a FNR of 7.0\% over the entire BE-SEL dataset, which is the more practical case because one would use all data available in practice.

Therefore, it is fair to say that VulDeePecker substantially outperforms two state-of-the-art code similarity-based vulnerability detection systems,
simply because VulDeePecker incurs a FPR of 5.7\% and a FNR of 7.0\% over the entire dataset. Nevertheless, it is important to note that deep learning-based vulnerability detection largely rely on the amount of data. This leads to:

\begin{insight}
VulDeePecker is more effective than code similarity-based vulnerability detection systems, which cannot detect vulnerabilities that are not caused by code clones and thus often incur high false negative rate. Nevertheless, the effectiveness of VulDeePecker is sensitive to the amount of data, which appears to be inherent to the nature of deep learning.
\end{insight}

\noindent{\bf Using VulDeePecker in practice.}
In order to further show the usefulness of VulDeePecker, we collected 20 versions of 3 software products: Xen, Seamonkey, and Libav.
These products are different from the target programs mentioned above.
We use VulDeePecker and the other vulnerability detection systems to detect the vulnerabilities in those software products.
As highlighted in Table \ref{Vulnerabilities detected by VulDeePecker},
VulDeePecker detected 4 vulnerabilities that have {\em not} been published in the NVD.
We manually checked and confirmed these vulnerabilities, and found that they have been published for other products and have been ``silently'' patched by the product vendors in the subsequent versions.
In contrast, these vulnerabilities are missed by almost all of the other vulnerability detection systems mentioned above, except that
Flawfinder detects the vulnerability corresponding to CVE-2015-4517 while missing the other three.

\section{Limitations}
\label{sec:Limitations}
The present design, implementation, and evaluation of VulDeePecker have several limitations, which suggest interesting open problems for future research.
First, the present design of VulDeePecker is limited to dealing with vulnerability detection by assuming source code of programs is available.
The detection of vulnerabilities in executables is a different and more challenging problem.

Second, the present design of VulDeePecker only deals with C/C++ programs. Future research needs to be conducted to adapt it to deal with other kinds of programming languages.

Third, the present design of VulDeePecker only deals with vulnerabilities related to library/API function calls. We will investigate how to detect the other kinds of vulnerabilities by leveraging the other kinds of {\em key points} mentioned above.

Fourth, the present design of VulDeePecker only accommodates {\em data flow} analysis (i.e., data dependency), but not {\em control flow} analysis (i.e., control dependency), despite the fact that the notion of code gadgets can accommodate data dependency and control dependency.
It is an important future work to improve the leverage of data flow analysis and accommodate control flow analysis to enhance vulnerability detection capabilities.

Fifth, the present design of VulDeePecker uses some heuristics
in labeling the ground truth of code gadgets, transforming code gadgets into their symbolic representations, transforming variable-length vector representations of code gadgets into fixed-length vectors. While intuitive, further research needs to be conducted to characterize the impact of these heuristics on the effectiveness of VulDeePecker.

Sixth, the present implementation of VulDeePecker is limited to the BLSTM neural network. We plan to conduct systematic experiments with other kinds of neural networks that could be used for vulnerability detection.

Seventh, the present evaluation of VulDeePecker is limited because the dataset only contains buffer error vulnerabilities and resource management error vulnerabilities.
We will conduct experiments on all available types of vulnerabilities.
Although we further tested VulDeePecker against 3 software products (i.e., Xen, Seamonkey, and Libav) and found 4 vulnerabilities that were not reported in the NVD and were ``silently''
patched by the vendors when releasing later versions of these products, these vulnerabilities were known rather than 0-day ones.
Extensive experiments need to be conducted against more software products to check whether VulDeePecker has the capability in detecting 0-day vulnerabilities.
In principle, this is possible because VulDeePecker uses pattern-based approach.

\section{Related Work}
\label{sec:Related_work}
We classify the related prior work into two categories: vulnerability detection (in relation to the purpose of the present paper), and program analysis (in the relation to the means for vulnerability detection).

\subsection{Prior work in vulnerability detection}

\noindent{\bf Pattern-based approach.}
This approach can be further divided to three categories.
In the first category,
patterns are generated {\em manually} by human experts
(e.g., open source tools Flawfinder \cite{FlawFinder}, RATS \cite{RATS}, and ITS4 \cite{ITS4},  commercial tools Checkmarx \cite{Checkmarx},
Fortify \cite{HP_Fortify}, and Coverity \cite{Coverity}). 
These tools often have high false positive rate or false negative rate. 
In the second category,
patterns are generated {\em semi-auto\-matically} from pre-classified vulnerabilities (
e.g.,
missing check vulnerabilities \cite{yamaguchi2013chucky},
taint-style vulnerabilities \cite{yamaguchi2015automatic}, and
information leakage vulnerabilities \cite{backes2009automatic}) and a pattern is specific to a type of vulnerabilities.
In the third category,
patterns are generated {\em semi-automa\-tically} from type-agnostic vulnerabilities (i.e., no need to pre-classify them into different types).
These methods use machine learning techniques, which rely on human experts for defining features to characterize vulnerabilities \cite{neuhaus2007predicting,shin2011evaluating,neuhaus2009beauty,yamaguchi2011vulnerability,yamaguchi2012generalized,grieco2016toward}.
Moreover, these methods cannot pin down the precise locations of vulnerabilities because programs are represented in coarse-grained granularity
(e.g., program \cite{grieco2016toward}, package \cite{neuhaus2009beauty}, component \cite{neuhaus2007predicting,scandariato2014predicting}, file \cite{moshtari2016evaluating,shin2011evaluating}, and function \cite{yamaguchi2011vulnerability,yamaguchi2012generalized}).

VulDeePecker falls into the pattern-based approach to vulnerability detection. In contrast to the studies reviewed above, VulDeePecker has two advantages.
First, it does {\em not} need human experts to define features for distinguishing vulnerable code and
non-vulnerable code. Second, it uses a {\em fine-grained} granularity to represent programs, and therefore can pin down the precise locations of vulnerabilities.

\noindent{\bf Code similarity-based approach.}
This approach has three steps.
The first step is to divide a program into some {\em code fragments}
\cite{jang2012redebug,kim2017vuddy,sajnani2016sourcerercc,pham2010detection,li2012cbcd}.
The second step is to represent each code fragment in the abstract fashion, including tokens \cite{jang2012redebug,kim2017vuddy,sajnani2016sourcerercc}, trees \cite{pham2010detection,jiang2007deckard}, and graphs \cite{pham2010detection,li2012cbcd}.
The third step is to compute the similarity between code fragments via their abstract representations obtained in the second step.

Compared with any pattern-based approach to vulnerability detection (including VulDeePacker), the code similarity-based approach has the advantage that a single instance of vulnerable code is sufficient for detecting the same vulnerability in target programs.
But it can only detect vulnerabilities in the Type I and Type II code clones \cite{rattan2013software}
(i.e., identical or almost-identical code clones), and some Type III code clones \cite{rattan2013software} (e.g., deletion, insertion, and rearrangement of statements).
In order to achieve a higher effectiveness of vulnerability detection,
human experts need to define features in order to automatically select the right code similarity algorithms for different kinds of vulnerabilities \cite{li2016vulpecker}.
However, even the enhanced approach with expert-defined features \cite{li2016vulpecker} cannot detect vulnerabilities that are {\em not} caused by code clones.
In contrast, VulDeePecker can detect vulnerabilities that may or may {\em not} caused by code clones, in an automatic fashion  (i.e., no need of human expert to define features).

\subsection{Prior work related to using deep learning for program analysis}
To the best of our knowledge, we are the first to use deep learning to detect software vulnerabilities, as inspired by the success of
deep learning in image processing, speech recognition, and natural language processing \cite{krizhevsky2012imagenet,hinton2012deep,pennington2014glove}.
In order to use deep learning for detecting software vulnerabilities, programs need to be represented in vectors.
There are two approaches for this purpose.
One is to map the {\em tokens} extracted from programs, such as data types, variable names, function names, and keywords, to vectors \cite{white2015toward};
the other is to map the {\em nodes} of abstract syntax trees extracted from programs, such as function definitions, function invocations, identifier declarations, and control flow nodes, to vectors \cite{mou2014building,wang2016automatically}.
VulDeePecker maps the tokens extracted from code gadgets to vectors,
while taking it into consideration that the lines of code in the code gadget is not necessarily consecutive.

Somewhat related work is the use of deep learning for software {\em defect} prediction \cite{wang2016automatically,yang2015deep}.
However, software defects are different from software vulnerabilities (i.e., methods for detecting defects cannot be used for detecting vulnerabilities in general)
\cite{morrison2015challenges}, and the file-level representation of programs in \cite{wang2016automatically} is too coarse-grained to pin down the locations of vulnerabilities.
Moreover, the defect prediction method presented in \cite{yang2015deep} is geared towards {\em code changes} rather than target programs as a whole.
Remotely related work is the use of deep learning for
purposes, like software language modeling \cite{white2015toward}, code cloning detection \cite{white2016deep}, API learning \cite{gu2016deep}, binary function boundary recognition \cite{shin2015recognizing},
and malicious URLs, file paths detection and registry keys detection \cite{saxe2017expose}.

\section{Conclusion}
\label{sec:Conclusion}

We have presented VulDeePecker, the first deep learning-based vulnerability detection system,
which aims to relieve human experts from the tedious and subjective work of manually defining features
and reduce the false negatives that are incurred by other vulnerability detection systems.
Since deep learning is invented for applications that are very different from vulnerability detection,
we have presented some preliminary principles for guiding the practice of applying deep learning to vulnerability detection.
These principles should be further refined because deep learning has great potential in solving the problem of vulnerability detection.
We have collected, and made publicly available, a useful dataset for evaluating the effectiveness of VulDeePecker and other deep learning-based vulnerability detection systems that will
be developed in the future.
Systematic experiments show that VulDeePecker can achieve much lower false negative rate than other vulnerability detection systems, while relieving human experts from the tedious work of manually defining features. For the 3 software products we experimented with (i.e., Xen, Seamonkey, and Libav), VulDeePecker detected 4 vulnerabilities,
which were not reported in the NVD and were ``silently'' patched by the vendors when they released later versions of these products.
In contrast, the other detection systems missed almost all of these vulnerabilities, except that one system detected 1 of these vulnerabilities and missed the other three vulnerabilities.

Open problems for future research are abundant, including the limitations of the present study discussed in Section \ref{sec:Limitations}. In particular, precisely characterizing the capabilities and limitations of deep learning-based vulnerability detection is an exciting research problem.

\section*{Acknowledgment}
We thank the anonymous reviewers for their comments that helped us improve the paper, and Marcus Pendleton for proofreading the paper.
This paper is supported by the National Basic Research Program of China (973 Program) under grant No.2014CB340600, the National Science Foundation of China under grant No. 61672249, the Shenzhen Fundamental Research Program under grant No. JCYJ20170413114215614, and the Natural Science Foundation of Hebei Province under grant No. F2015201089.
Shouhuai Xu is supported in part by NSF Grant \#1111925 and ARO Grant \#W911NF-17-1-0566.
Any opinions, findings, conclusions or recommendations expressed in this material are those of the authors and do not reflect the views of the funding agencies.

\bibliographystyle{IEEEtranS}
\bibliography{ndss-sample}

\appendix
\renewcommand{\appendixname}{Appendix~\Alph{section}}
\subsection{LSTM cells}
\label{sec:appendix-LSTM-cells}

The BLSTM layers in the BLSTM neural network contain a complex structure called {\em LSTM cells}, which are briefly reviewed below and referred to \cite{hochreiter1997long} for greater details.

Let $\odot$ denote the element-wise multiplication, $tanh$ denote the hyperbolic tangent function $\frac{exp(x) - exp(-x)}{exp(x)+exp(-x)}$, and $\sigma$ denote the sigmoid function $\frac{1}{1+exp(-x)}$.

Each LSTM cell, denoted by $c$, uses an input gate $i$ (i.e.,
the input data),  a forget gate $f$ (i.e., the state flow of the cell), and an output gate $o$ (i.e., the output of module) to control the data flow through the neural network.
The output $h^l_t$ of the layer $l$ at the time $t$ is:
\begin{equation*}
  h^l_t = o^l_t \odot tanh(c^l_t),
\end{equation*}
where the output gate $o^l_t$ of the layer $l$ at the time $t$ is:
\begin{equation*}
  o^l_t = \sigma(\bm{W^l_{xo}x^l_t} + \bm{W^l_{ho}h^l_{t-1}} + \bm{W^l_{co}c^l_t} + \bm{b^l_o})
\end{equation*}
and the state of LSTM cell $c^l_t$ of the layer $l$ at the time $t$ is:
\begin{equation*}
  c^l_t = f^l_t \odot c^l_{t-1} + i^l_t \odot tanh(\bm{W^l_{xc}x^l_t} + \bm{W^l_{hc}h^l_{t-1}} + \bm{b^l_c}).
\end{equation*}

The forget gate $f^l_t$ and the input gate $i^l_t$ of the layer $l$ at the time $t$ are calculated as follows:
\begin{eqnarray*}
  f^l_t &=& \sigma(\bm{W^l_{xf}x^l_t} + \bm{W^l_{hf}h^l_{t-1}} + \bm{W^l_{cf}c^l_{t-1}} + \bm{b^l_f}), \\
  i^l_t &=& \sigma(\bm{W^l_{xi}x^l_t} + \bm{W^l_{hi}h^l_{t-1}} + \bm{W^l_{ci}c^l_{t-1}} + \bm{b^l_i}),
\end{eqnarray*}
where $\bm{x^l_t}$ is the input to layer $l-1$ ($l > 1$) or the input of the network ($l = 1$),
$\bm{W^l_{xi}}$, $\bm{W^l_{xf}}$, $\bm{W^l_{xo}}$, $\bm{W^l_{xc}}$
are the weight matrices connecting $\bm{x^l_t}$ with the input gate, the forget gate, the output gate, and the LSTM cell input,
$\bm{W^l_{hi}}$, $\bm{W^l_{hf}}$, $\bm{W^l_{ho}}$, $\bm{W^l_{hc}}$
are the weight matrices connecting $\bm{h^l_{t-1}}$ with the input gate, the forget gate, the output gate, and the LSTM cell input, and
$\bm{b^l_i}$, $\bm{b^l_f}$, $\bm{b^l_o}$, $\bm{b^l_c}$ are the bias items of the input gate, the forget gate, the output gate, and the LSTM cell input.

\subsection{Library/API function calls selected by Checkmarx}
\label{sec:appendix:library-and-API-functions}

Table \ref{Table_sensitive_function} summarizes the C/C++ library/API function calls related to the two types of vulnerabilities, buffer error (CWE-119) and resource management error (CWE-399), where ``*'' represents the wildcard.
These library/API function calls are generated by the commercial product Checkmarx \cite{Checkmarx}.

\begin{table}[!htbp]
\renewcommand{\arraystretch}{1.2}
\caption{Library/API function calls related to two types of vulnerabilities}
\label{Table_sensitive_function}
\centering
\begin{tabular}{|c|p{0.38\textwidth}|}
\hline
CWE ID & C/C++ library/API function calls related to vulnerabilities \\
\hline
CWE-119 & cin,
getenv, getenv\_s, \_wgetenv, \_wgetenv\_s, catgets, gets, getchar, getc, getch, getche, kbhit, stdin,
getdlgtext, getpass, scanf, fscanf, vscanf, vfscanf,
istream.get, istream.getline, istream.peek, istream.read*, istream.putback, streambuf.sbumpc, streambuf.sgetc, streambuf.sgetn, streambuf.snextc, streambuf.sputbackc, SendMessage, SendMessageCallback, SendNotifyMessage, PostMessage, PostThreadMessage, recv, recvfrom, Receive, ReceiveFrom, ReceiveFromEx, Socket.Receive*,
memcpy, wmemcpy, \_memccpy, memmove, wmemmove, memset, wmemset, memcmp, wmemcmp, memchr, wmemchr, strncpy, \_strncpy*, lstrcpyn, \_tcsncpy*, \_mbsnbcpy*, \_wcsncpy*, wcsncpy, strncat, \_strncat*, \_mbsncat*, wcsncat*, bcopy, strcpy, lstrcpy, wcscpy, \_tcscpy, \_mbscpy, CopyMemory, strcat, lstrcat, lstrlen, strchr, strcmp, strcoll, strcspn, strerror, strlen, strpbrk, strrchr, strspn, strstr, strtok, strxfrm, readlink, fgets, sscanf, swscanf, sscanf\_s, swscanf\_s, printf, vprintf, swprintf, vsprintf, asprintf, vasprintf, fprintf, sprint, snprintf, \_snprintf*, \_snwprintf*, vsnprintf, CString.Format, CString.FormatV, CString.FormatMessage, CStringT.Format, CStringT.FormatV, CStringT.FormatMessage, CStringT.FormatMessageV, syslog, malloc,
Winmain, GetRawInput*, GetComboBoxInfo, GetWindowText, GetKeyNameText, Dde*, GetFileMUI*, GetLocaleInfo*, GetString*, GetCursor*, GetScroll*, GetDlgItem*, GetMenuItem*\\
\hline
CWE-399 & free, delete, new, malloc, realloc, calloc, \_alloca, strdup, asprintf, vsprintf, vasprintf, sprintf, snprintf, \_snprintf, \_snwprintf, vsnprintf\\
\hline
\end{tabular}
\end{table}

\end{document}